\documentclass[amsmath,amstex,eqsecnum,preprint,pre,aps]{revtex4}

\usepackage{amsmath}
\usepackage{graphicx}
\usepackage{bm}
\usepackage{float}

\begin{document}

\title{ Generalized Langevin Equation for Tracer Diffusion in Atomic Liquids}
\author{ Patricia Mendoza-M\'endez, Leticia L\'opez-Flores,}

\address{Facultad de Ciencias Fisico-Matem\'aticas,
Benem\'{e}rita Universidad Aut\'{o}noma de Puebla, C.P.72000, Puebla, Pue., M\'{e}xico}

\author{ Alejandro
Vizcarra-Rend\'on,}

\address{Unidad Acad\'emica de F\'isica, Universidad Aut\'onoma de Zacatecas,
Paseo la Bufa y Calzada Solidaridad, 98600, Zacatecas, Zac., Mexico}

\author{ Luis E. S\'anchez-D\'iaz,  and Magdaleno Medina-Noyola}
\address{Instituto de F\'{\i}sica {\sl ``Manuel Sandoval Vallarta"},
Universidad Aut\'{o}noma de San Luis Potos\'{\i}, \'{A}lvaro
Obreg\'{o}n 64, 78000 San Luis Potos\'{\i}, SLP, M\'{e}xico}

\date{\today}

\begin{abstract}

We derive the time-evolution equation that describes the Brownian motion of labeled individual tracer particles in a simple model atomic liquid (i.e., a system of $N$  particles whose motion is governed by Newton's second law, and interacting through spherically symmetric pairwise potentials). We base our derivation on the generalized Langevin equation formalism, and find that the resulting time evolution equation is formally identical to the generalized Langevin equation that describes the Brownian motion of individual tracer particles in a colloidal suspension in the absence of hydrodynamic interactions. This formal dynamic equivalence implies the long-time indistinguishability of some dynamic properties of both systems, such as their mean squared displacement, upon a well-defined time scaling. This prediction is tested here by comparing the results of molecular and Brownian dynamics simulations performed on the hard sphere system.

\end{abstract}

\pacs{61.20.Lc, 82.70.Dd}

\maketitle

\section{Introduction.}\label{sectionI}

It is well known that under some circumstances the  phenomenology of atomic liquids finds an almost perfect correspondence in the phenomenology of colloidal fluids \cite{pusey0,nagele0,deschepperpusey1, deschepperpusey2}. This seems to be particularly true regarding the rather complex dynamic behavior of these systems as they approach the glass transition \cite{lowenhansenroux,szamelflenner,puertasaging}.
Although it is clear that this analogy has some fundamental
limitations (such as the presence of many-body hydrodynamic
interactions in colloidal systems), one can be confident, for example, that the phase behavior of colloidal and atomic systems with identical interaction potentials will, of course, be the same. Thus, if one approaches this problem with a dynamic
simulation technique, one is confident that the equilibrium phase
diagram of a specific system (say a Lennard-Jones liquid) will be
independent of the simulation technique employed in its
determination (either molecular or Brownian dynamics)
\cite{tildesley}. Time-dependent and dynamic properties, on the other hand,
are expected in general to depend on the specific microscopic
transport mechanisms. Nevertheless, some features associated with the collective,
long-time behavior of the system also seem to be rather insensitive
to the microscopic short-time dynamics. For example, it has been
suspected, and partially corroborated, that for a given model system
(i.e., same pair potential) standard molecular dynamics will lead to
essentially the same dynamic arrest scenario as Brownian dynamics
\cite{lowenhansenroux,szamelflenner,puertasaging}. Determining the range of validity of this dynamic analogy, however, continues to be a relevant topic in the study of the dynamics of liquids.

From the theoretical side, for example, one would like to have a
unified description of the macroscopic dynamics of both, colloidal
and atomic liquids, which explicitly exhibits the origin of the
similarities and differences in their macroscopic dynamics. This topic has been
addressed in the framework of the mode coupling theory of the ideal
glass transition \cite{goetze1}, originally developed for Newtonian
liquids, but also adapted to Brownian systems. Such attention was
focussed on the similarity of the long-time dynamics of Newtonian
and Brownian systems in the neighborhood of the glass transition
\cite{szamellowen}. A number of issues, however, remain open
\cite{szamelflenner}, one important question referring to the validity and limitations
of this long-time similarity under general conditions, such as those
involving ordinary thermodynamically stable fluids, and not
necessarily associated with the glass transition. In this sense, one possible general framework for such theoretical analysis is the concept of the generalized Langevin equation (GLE) \cite{delrio,faraday}.

The GLE formalism describes the dynamics of the thermal fluctuations $\delta a_{i}(t)\ (\equiv a_{i}(t)-a^{eq}_{i})$ of the instantaneous value of the macroscopic variables $ a_{i}(t)$ ($i=1,2,...,\nu$), around its equilibrium value $a^{eq}_{i}$, and has the structure of the most general linear stochastic equation with additive noise for the vector $\delta \mathbf{a}(t)=\left[\delta a_{1}(t),\delta a_{2}(t),...,\delta a_{\nu }(t)\right]^{\dagger} $ (with the dagger indicating transpose). The GLE equation has been widely used in the description of thermal fluctuation phenomena in simple liquid systems, and Boon and Yip's textbook \cite{boonyip} contains a detailed account of its early use to describe the dynamics of simple liquids. Although this stochastic equation is conventionally associated with the Mori-Zwanzig projection operator formalism \cite{zwanzig,mori}, in reality its structure is not a consequence of the hamiltonian basis of Mori-Zwanzig's derivation; instead, it is essentially equivalent to the mathematical condition of stationarity \cite{delrio}.

Understood in the latter manner, the GLE formalism was first employed in Ref. \cite{faraday} to derive the equation of motion of an individual tracer particle in a colloidal suspension without hydrodynamic interactions. Such an equation reads
\begin{equation}
M{\frac{d{\bf v}(t)}{dt}}=  -\zeta ^{(s)}{\bf v}(t)+{\bf f}
 ^{(s)}(t)- \int_0^t dt' \Delta
\zeta(t-t')  {\bf v}(t')+ {\bf F} (t), \label{gletracerdif}
\end{equation}
where $M$ is the mass and ${\bf v}(t)$ the
velocity of the tracer particle, while $\zeta^{(s)}$ is the friction coefficient
caused by the frictional resistance of the supporting solvent and ${\bf f}^{(s)}(t)$ the associated random force. The memory term involving the time-dependent friction function $\Delta \zeta(t)$, and its associated random force ${\bf F}(t)$, are the friction and fluctuating forces that originate in the time-evolution of the cage of surrounding colloidal particles. Under well defined approximations, the exact result for the time-dependent friction function $\Delta \zeta (t)$ derived in Ref. \cite{faraday} was shown there to reduce to the following approximate expression in terms of the collective and self intermediate scattering functions (ISFs) $F(k,t)$ and $F_S(k,t)$,
\begin{equation}
\Delta \zeta (t) =\frac{k_BT}{3\left( 2\pi \right) ^{3}n}\int d
{\bf k}\left[\frac{ k[S(k)-1]}{S(k)}\right] ^{2}F(k,t)F_S(k,t).
\label{dzdt0}
\end{equation}
In this equation $T$ is the temperature, $n$ the number concentration, and $S(k)$ the static structure factor of the bulk suspension. This result, together with similarly general expressions for  $F(k,t)$ and $F_S(k,t)$ also derived within the GLE formalism \cite{scgle0}, was later employed in the construction of the self-consistent generalized Langevin equation (SCGLE) theory of colloid dynamics \cite{scgle1,scgle2}, eventually applied to the description of dynamic arrest phenomena \cite{rmf,todos1,todos2}, and more recently, to the construction of a first-principles theory of equilibration and aging of colloidal glass-forming liquids \cite{noneqscgle0,noneqscgle1}.

With the aim of investigating the relationship between the dynamics of atomic and Brownian liquids, in this work we apply the GLE formalism to derive the generalized Langevin equation that describes the motion  of individual tracer particles in simple \emph{atomic } liquids, thus extending to these systems the results of Ref. \cite{faraday} reviewed above. The most remarkable prediction of the derivation presented here is that the resulting stochastic equation for the velocity ${\bf v}(t)$ of the atomic tracer turns out to be formally identical to the colloidal case described by the two equations above, with the solvent friction coefficient $\zeta ^{(s)}$ replaced by a kinetic (or `Doppler') friction coefficient $\zeta ^0$ determined by kinetic-theoretical arguments.

Since the concept of kinetic friction may involve a rather subtle use of otherwise simple and well established concepts, we start this paper in section II by providing a simple and intuitive
description of the short-time random motion performed by an individual tracer particle in an
\emph{atomic} liquid, as a consequence of molecular collisions. Such description exhibits the fact that the resulting random motion must be described by the same stochastic mathematical
model that describes the Brownian motion of a tracer particle in a
colloidal fluid. This is just the mathematical model underlying the ordinary Langevin equation (Eq. (\ref{gletracerdif}) above without the time-dependent friction term and its associated random force). Thus, also in the atomic case, a relaxation time $\tau_0$ of the velocity, due to
the friction force $-\zeta^0{\bf v}(t)$, defines the crossover
from ballistic  to diffusive  motion. The fundamental difference
lies in the physical origin of the friction force  $-\zeta^0{\bf
v}(t)$ and in the determination of the friction coefficient
$\zeta^0$. In a Brownian liquid the friction force $-\zeta ^{(s)}{\bf v}(t)$ is caused by the
supporting solvent; hence, $\zeta^{(s)}$ assumes its Stokes value. In contrast, as discussed  in section II, in
a Newtonian liquid the friction force  $-\zeta^0{\bf
v}(t)$ is not caused by any external material agent but by the
unimpeded tendency to establish or restore, through molecular
collisions, the equipartition of the energy available for
distribution  among  the kinetic energy degrees of freedom of the
system. Thus, the corresponding value of $\zeta^0$ that emerges
from these considerations is provided by Einstein's relation, with
a kinetic-theoretically determined diffusion coefficient.

The discussion of the Langevin equation for atomic liquids is continued in sections \ref{sectionIII} and \ref{sectionIV}, where a more formal and complete derivation is provided. Thus, in section \ref{sectionIII}
we show that the force on the tracer particle can be written as an integral of the divergence of the stress tensor $\stackrel {\leftrightarrow}{\Pi} (\mathbf{r},t)$, whose kinetic component $\stackrel {\leftrightarrow}{\Pi}_K (\mathbf{r},t)$ is the origin of the Doppler friction force  $-\zeta^0{\bf v}(t)$ and its associated random force, whereas the term involving the configurational component $\stackrel {\leftrightarrow}{\Pi}_U (\mathbf{r},t)$ describes the effects of
the ordinary conservative direct forces (electrostatic, van der Waals, etc,) exerted by the surrounding particles. It is also seen that the latter effects enter additively in the Langevin equation of the atomic liquid, as an  integral term that is linear in the instantaneous local density of the surrounding particles. In section \ref{sectionIV} we demonstrate that this linear coupling of the motion of the tracer with the local density of the surrounding particles leads to a time-dependent configurational friction term and to its corresponding random force, thus resulting in a generalized langevin equation for the velocity of a tracer particle, which turns out to be formally identical to its colloidal counterpart in Eq. (\ref{gletracerdif}).

The main predictions of the resulting generalized Langevin equation for atomic liquids are then discussed in Section \ref{sectionV}. These include well defined scaling rules that exhibit the identity between the long-time dynamics of atomic and colloidal liquids. There we test these scalings by comparing the simulation results for a given model system (the hard sphere fluid) using both, molecular dynamics and Brownian dynamics simulations. The last section summarizes the most relevant conclusions, and discusses some limitations and potential applications of the results of the paper.

\section{Ballistic and diffusive regimes in a simple atomic
liquid.} \label{sectionII}

Let us consider a simple atomic fluid, formed by $N$ spherical
particles in a volume $V$ whose microscopic dynamics is described by
Newton's equations,
\begin{equation}
M{\frac{d{\bf v}_{i}(t)}{dt}}= \sum_{j\neq i}{\bf F}_{ij}(t),\quad
(i=1,2,\ldots ,N), \label{eq1}
\end{equation}
where $M$ is the mass and ${\bf v}_{i}(t)=d{\bf r}_{i}(t)/dt$ the velocity of the $i$th
particle at position ${\bf r}_{i}(t)$, and in which the interactions between the particles are
represented by a sum of pairwise forces, with ${\bf F}_{ij}=-\nabla_i u(|{\bf r}_{i}-{\bf r}_{j}|)$ being the force exerted on particle $i$ by particle $j$. Our general aim is to establish a connection between the microscopic dynamics described by these (Newton's) equations, and the equation that describes the random motion of any representative individual tracer particle of the liquid.

In this section we discuss a simple and intuitive (albeit possibly subtle) physical picture, which plays a central role in our theoretical effort to establish such connection. We start by recalling the physical meaning of two fundamental concepts in the dynamics of an atomic fluid, namely, the mean free time, $\tau_0$, and the mean free path, $l_{0}$, which represent the characteristic time- and length-scales at which the crossover from the short-time ballistic motion of the atoms to their long-time diffusive transport occur. It is well known \cite{hansenmcdonald,boonyip} that for correlation times $t$ much shorter than $\tau_0$, and for distances much shorter than  $l_{0}$,  all the particles move ballistically, so that for $t\ll\tau_0$, the mean squared displacement (MSD) is given by $ <(\Delta \textbf{r}(t))^2> \ \approx \ 3v_0^2 t^2$, with
$v_0\equiv (k_BT/M)^{\frac{1}{2}}$ being the thermal velocity.

For times $t$ much longer than $\tau_0$, each particle has undergone
many collisions, and its motion can be represented as a sequence of
(ballistic) random flights of mean length $l_{0}$ and mean flight-time $\tau_0$, traveled at a random velocity that has zero mean and covariance $v_0^2$. From the theory of random flights, however, it is well-known \cite{chandrasekhar} that the motion represented by such a sequence of random displacements, will become diffusive in the long-time limit. This means that it will be characterized by a mean squared displacement that, for $t\gg \tau_0$, will increase linearly with time, $ <(\Delta \textbf{r}(t))^2> \ \approx  6D^0t$. Furthermore, the corresponding  diffusion coefficient $D^0$ will be given by $D^0 = (l_0)^2/\tau_0$.

Thus, we conclude that the MSD of a representative tracer particle will exhibit two well-defined limiting behaviors in two opposite time regimes, namely, it will be ballistic at short times, $ <(\Delta \textbf{r}(t))^2> \ \approx \ 3v_0^2 t^2$ for $t\ll \tau_0$, and diffusive at long times, $ <(\Delta \textbf{r}(t))^2> \ \approx  6D^0t$ for $t \gg \tau_0$. The simplest  mathematical model that provides a full description of the crossover of $<(\Delta \textbf{r}(t))^2>$ from the first to the second of these two exact limits, is provided by a Gaussian stationary stochastic process, described by a linear stochastic equation with additive noise for the instantaneous velocity ${\bf v}(t)$ of the tracer particle, i.e., by \cite{keizer}
\begin{equation}
M{\frac{d{\bf v}(t)}{dt}}= -\zeta^0 {\bf v}(t)+{\bf f}^0 (t),
\label{langevin0}
\end{equation}
with ${\bf f}^0 (t)$ being a ``purely random" (or ``white") noise, i.e., a stationary  and Gaussian stochastic process with zero mean ($\overline{{\bf f}^0(t)}=\textbf{0}$), uncorrelated with
the initial value ${\bf v}_0$ of the velocity fluctuations, and
delta-correlated with itself.  In fact, the stationarity condition is
in reality equivalent to the fluctuation-dissipation relation
between the random and the dissipative terms in Eq.
(\ref{langevin0}), $\overline{{\bf f}^0(t){\bf f}^0(t')}=\langle
\textbf{v}_0 \textbf{v}_0\rangle M \zeta^0 2\delta(t-t')$, where $\langle
\textbf{v}_0\textbf{v}_0\rangle$ is the stationary covariance of the initial velocities. Within the additional physical assumption that identifies  the stationary state described by Eq. (\ref{langevin0}) with the thermodynamic equilibrium state, we have that this covariance is determined by the equipartition theorem, $\langle
\textbf{v}_0\textbf{v}_0\rangle= (k_BT/M)\stackrel{\leftrightarrow }{{\bf I}}$ (with
$\stackrel{\leftrightarrow }{{\bf I}}$ being the 3$\times$3 cartesian unit tensor).

Clearly, this equation is formally identical to the ordinary Langevin equation \cite{langevin} for the instantaneous velocity ${\bf v}(t)$ of a colloidal particle in a solvent, in which case, Stoke's result $\zeta ^{(s)}=3\pi \eta \sigma$ (with $\eta$ being the viscosity of the solvent and $\sigma$ the diameter of the colloidal particle) provides an independent determination of the friction coefficient $\zeta ^{(s)}$  \cite{mcquarrie}. In the present case, however, there is no supporting solvent to produce friction, and hence, identifying the origin and determining the value of $\zeta^0$ requires slightly more subtle arguments. The simplest manner to describe its physical origin may be found in Uhlenbeck and Ornstein's brief reference to the so-called \emph{Doppler} friction \cite{ornsteinuhlenbeck}. These authors point out that any tracer particle colliding with the particles of a gas, whose size $\sigma$ is smaller than the mean free path $l_0$, will be subjected to Doppler friction, caused by the fact that ``when the tracer particle is moving, say to the right, will be hit by more molecules from the right than from the left". In the following section we shall provide a more formal derivation of this kinetic friction effect. At this point, however, we provide simple arguments for its quantitative determination.

To determine $\zeta^0$ in the present case we first notice that the mathematical solution of Eq. (\ref{langevin0}) for the MSD is such that at long times, $ <(\Delta \textbf{r}(t))^2> \ \approx \ 6  (k_BT/ \zeta^0) t $. This determines a  self-diffusion coefficient $D^0$ in terms of  $\zeta^0$ through Einstein's relation, $D^0 =  k_BT/\zeta^0$. Thus, we must only determine either $\zeta^0$ or $D^0$. In our case, we write Einstein's relation as
\begin{equation}
\zeta^0\equiv k_BT/D^0,
\label{einstein}
\end{equation}
and determine $D^0$ independently, borrowing the arguments developed in the elementary kinetic theory of gases \cite{mcquarrie}. For this, we recall that  $D^0 = (l_0)^2/\tau_0$, which, since $l_0/\tau_0 =v_0$, can be written as $D^0 = v_0l_0$. We then estimate the mean free path  $l_{0}$ to be given by $l_{0} \sim
1/n\sigma^2$, with $n\equiv N/V$ and with $\sigma$ being the
collision diameter of the particles. Thus, we must have that $D^0
\sim \sqrt{k_BT/M}/(n\sigma^2)$. In fact, the rigorous value of
$D^0$ is  \cite{chapmancowling}
\begin{equation}
D^0\equiv \frac{3}{8\sqrt
\pi}\left(\frac{k_BT}{M}\right)^{1/2}\frac{1}{n\sigma^2}.
\label{dkinetictheory}
\end{equation}
This expression, together with Einstein's relation above, determines the value of the kinetic friction coefficient $\zeta^0$ of an atomic fluid.

Let us now discuss some of the implications of the \emph{atomic} Langevin equation defined by Eqs. (\ref{langevin0}), (\ref{einstein}), and (\ref{dkinetictheory}). Since the Langevin equation itself is mathematically identical to the ordinary (i.e., colloidal) Langevin equation, its solution is also formally the same. For example, from Eq. (\ref{langevin0}), and the assumed properties of ${\bf f}^0(t)$ one can evaluate the velocity auto-correlation function (VAF)
\begin{equation}
V(t)\ \equiv <\textbf{v}(t)\cdot \textbf{v}(0)>/3,
\end{equation}
with the result
\begin{equation}
V(t)  = v_0^2 e^{-t/\tau_S}, \label{vdt0}
\end{equation}
with
\begin{equation}
\tau_S \equiv M/\zeta^0
\label{taub}
\end{equation}
being the velocity relaxation time.

The MSD, normalized as
\begin{equation}
W(t)\
\equiv <(\Delta \textbf{r}(t))^2>/6,
\label{msdaswdt}
\end{equation}
is related with the VAF
by means of the exact relationship
\begin{equation}
W(t)=\int_0^t (t-t')V(t')dt' \label{wdtvdt}
\end{equation}
or, in terms of the Laplace transforms (LT) $W(z)$ and $V(z)$, as
\begin{equation} W(z)=V(z)/z^2.
\label{wdzvdz}
\end{equation}
This exact relationship can be written, for the particular form of the VAF in Eq. (\ref{vdt0}), as the following differential equation,
\begin{equation}
\tau_S\frac{dW(t)}{dt} +W(t) =D^0t,
\label{wdtintdifec0}
\end{equation}
whose solution  reads
\begin{equation}
W(t)   = D^0 \tau_S \left[
\frac{t}{\tau_S}-1+e^{-\frac{t}{\tau_S}} \right]. \label{wdt0}
\end{equation}
This expression interpolates $W(t)$ between its corresponding short- and long-time asymptotic limits,
\begin{equation}
W(t)     \approx \frac{1}{2} v_0^2 t^2, \mbox{ for $ t\ll \tau_0$} \label{wdt0short}
\end{equation}
and
\begin{equation}
W(t)  \approx D^0t, \mbox{ for  $t\gg \tau_0$}.
\label{wdt0long}
\end{equation}
In addition, it also exhibits the fact that the crossover from ballistic to diffusive motion is most naturally described using  $\tau_S$ as the unit of time.

Let us notice, however, that for atomic liquids the relaxation time $\tau_S$ is identical to the mean free time $\tau_0$, since $\tau_S=M/\zeta^0=(k_BT/\zeta^0)(M/k_BT)=D^0/v_0^{2}= (l_0^2/\tau_0)(\tau_0/l_0)^2=\tau_0$. Thus, for an atomic liquid the mean free time is the most natural time unit, and the mean free path $l_0=v_0\tau_0$ the most natural length unit since, for an atomic liquid, Eq. (\ref{wdt0}) can be rewritten in terms of the scaled time $t^*\equiv t/\tau_0$ and the scaled MSD $w(t^*) \equiv \ W(t)/l_0^2 $ as
\begin{equation}
w(t^*)   =  \left[t^*-1
+e^{-t^*} \right]. \label{wdtstar}
\end{equation}
To illustrate the validity of this result, in Fig. 1 we compare
it with the molecular dynamics
simulation data (solid circles) for $w(t^*)$ in a fluid of hard
spheres of diameter $\sigma$ at a small but finite volume fraction
$\phi\equiv \pi n \sigma^3/6 $, namely, at $\phi =0.1$. Clearly,
the scaled solution (\ref{wdtstar}) of the atomic Langevin equation (solid curve) lies very close to  the simulation data of the MSD.

\begin{figure}[ht]
\includegraphics[scale=.3]{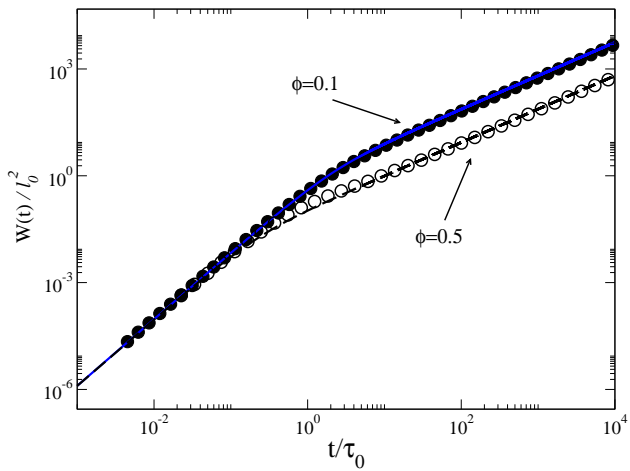}
\caption{Scaled mean squared displacement $[W(t)/l_0^2]$ as a function of the scaled time $ [ t/\tau_S]$, with $l_0$ and $\tau_S=\tau_0$ being the mean free path and mean free time, of a hard-sphere fluid at volume fractions $\phi=0.1$ (solid circles) and $\phi=0.5$ (empty circles) generated by molecular dynamics simulations of soft sphere systems using the velocity Verlet's algorithm \cite{tildesley} and the soft-to-hard-sphere dynamic correspondence of Ref. \cite{dynamicequivalence}. The solid line corresponds to the low density limit, Eq. (\ref{wdtstar}), and the dashed line corresponds to the expression in Eq. (\ref{wdtstar1}) with $D^*=0.099$. } \label{fig1}
\end{figure}

The dimensionless quantities $w(t^*)$ and $t^*$ serve to highlight an important scaling that derives directly from the atomic Langevin equation in Eq. (\ref{langevin0}). At the same time, however, these scaled variables hide the specificity of the actual magnitudes involved in concrete real physical situations. For example, let us think of a typical atomic liquid, such as argon ($M\approx 40$ amu, $\sigma \approx 3.8 \AA$), and rewrite the expression for $D^0$ in Eq. (\ref{dkinetictheory}) as
\begin{equation}
D^0\equiv \frac{\sqrt{\pi}}{16\phi}\left[\sigma
\left(\frac{k_BT}{M}\right)^{1/2}\right].
\label{dkinetictheory2}
\end{equation}
Then, for $T\approx 300 ^\circ K$  and $\phi\approx 0.1$, representative of a moderately dilute gas at room temperature, we find that  $D^0 \approx 1.0 \times 10^{-7} {\rm m}^2/{\rm s} $,  $l_0\approx 4.2 \AA$, and $\tau_S=\tau_0\approx 1.7$ps. Let us now compare this crossover timescale $\tau_S$ with the structural relaxation timescale $\tau_I\equiv d^2/D^0$, the time it would take any particle to {\it diffuse} one interparticle mean distance $d\equiv n^{-1/3}$ with the diffusion coefficient  $D^0$.  For the conditions under consideration we find that $\tau_I\approx4.16$ps, so that $\tau_I$ is only about twice longer than $\tau_S$. This means that there is nothing like a time-scale separation between the crossover from ballistic to diffusive motion and the crossover from free to collective relaxation (i.e., from the so-called $\beta$-processes to the slower $\alpha$-processes). Although this timescale degeneration disappears gradually as the density increases, giving rise to the full separation of time scales characteristic of the approach to the glass transition, this degeneration of time scales is a characteristic  feature of atomic dynamics throughout the whole stable liquid phase of these simple systems. To illustrate this, notice that taking $T\approx 83^\circ K$ and $\phi\approx 0.5$, representative of the freezing conditions of Argon, leads to $D^0 \approx 0.11 \times 10^{-7} {\rm m}^2/{\rm s}$,  $l_0\approx 0.84 \AA$, $\tau_S=\tau_0\approx 0.64$ps, and $\tau_I\approx13.5$ps., so that now $\tau_I\approx 20 \tau_S$.

This situation must now be compared with its corresponding colloidal analog. In this case, the velocity relaxation time $\tau_S=M/\zeta^0$ is no longer identical to the mean free time. Instead, it is determined by the mass M of the particle and by the solvent friction coefficient $\zeta ^{(s)}$, now given by its Stokes value $\zeta ^{(s)}= 3\pi \eta \sigma$, with $\eta$ being the viscosity of the solvent. Thus, consider a micron-sized colloidal particle in water at room temperature and with the same mass density as water itself, so that $\sigma \approx 1.0 \mu$, $M \approx 0.52\times 10^{-15} $ kg, and $\zeta ^{(s)} \approx 0.94 \times 10^{-8}$ kg/s (taking $\eta \approx 10^{-3}$ kg/m$\cdot$s for the viscosity of water). This leads to the estimate  $D^0 \approx 4.4 \times 10^{-13}{\rm m}^2/{\rm s}$ for the ($\phi$-independent) Einstein diffusion coefficient $D^0=k_BT/\zeta ^{(s)}$. The corresponding estimate of $\tau_S=M/\zeta ^{(s)}$ is, then, $\tau_S\approx 5.5\times 10^{-8}$s, whereas the value of  $\tau_I=d^2/D^0$ (for $\phi=0.1$) is $\tau_I\approx 6.87$s. We thus immediately notice a really dramatic separation of timescales, expressed by the fact that now $\tau_I \approx 10^{8}\tau_S$. As a consequence,
in a colloidal suspension, the crossover from ballistic to diffusive motion (occurring at very early times $t\approx \tau_S$) and the crossover from free to collective difusion (occurring at times $t\approx \tau_I$), are separated by nearly 8 decades. Thus, if one were interested in observing  the ballistic-to-diffusive crossover in our illustrative colloidal suspension, one would have to measure $W(t)$ in the time window corresponding to Fig. 1, i.e, for times centered in the regime $t\approx \tau_S \approx 10^{-8}$s. This crossover, however, will be completely shifted to unobservable, extremely short times, when probing only the structural relaxation of the colloidal liquid for times $t$ in the  milisecond range and above (i.e., $t\approx \tau_I$).  For this reason in the description of the dynamics of suspensions the inertial term of the Langevin equation is normally neglected, $M{\frac{d{\bf v}(t)}{dt}}\approx 0$, leading to the {\it overdamped} Langevin equation, $ -\zeta ^{(s)} {\frac{d{\bf r}(t)}{dt}}+{\bf f} ^{(s)} (t)=0$, whose solution for the MSD is $W(t)  \approx D^0t$. This, however, is just the diffusive limit in Eq. (\ref{wdt0long}).

The main conclusion of these illustrative estimates is, thus, that in colloidal liquids this wide separation of timescales is always  present, whereas it is virtually nonexistent in an atomic liquid. As mentioned above, however, this disparity between atomic and colloidal liquids will disappear in the metastable regime due to the dominant effect of  interparticle interactions in the determination of the structural relaxation processes. To introduce the discussion of these effects, in Fig. 1 we have also included the molecular dynamics simulation results for the MSD of the hard-sphere liquid  at freezing conditions, $\phi=0.5$ (empty circles). These results illustrate the deviations from the ideal behavior
described by Eq. (\ref{wdtstar}), which originate from the structural effects of the interparticle interactions (very moderate, and almost imperceptible in the results for $\phi=0.1$). The comparison between the results for the MSD at these two volume fractions clearly show that in both cases the ballistic short-time behavior (i.e., for times $t\le \tau_0$) is accurately described by Eq. (\ref{wdtstar}). Beyond the crossover  time $\tau_0$, however, the results for $\phi=0.1$ exhibit only the kinetic effects of the interparticle interactions, whereas the results for $\phi=0.5$ exhibit the combination of both, the kinetic and the structural effects of the interparticle forces.

One of the results of the following sections will be a simple expression for  $W(t)$ that preserves the short-time
limit $W(t)\approx \frac{1}{2} v_0^2 t^2$, but replaces the ideal
long-time limit $W(t)  \approx D^0t$ by the correct one, $W(t)
\approx D_L t,$ where $D_L$ is the long-time self-diffusion
coefficient  (see end of Sec. \ref{sectionIV}). This expression reads
\begin{equation}
w(t^*)   =  D^{*2}\left[t^*/D^*-1
+e^{-t^*/D^*} \right], \label{wdtstar1}
\end{equation}
with the parameter $D^* \equiv D_L/D^0$ being the ratio of the
{\it long-time} to the short-time self-diffusion coefficients. In fact, it is the solution of
\begin{equation}
\frac{dw(t^*)}{dt^*} + \frac{w(t^*)}{D^*} =t^*,
\label{wdtintdifec1}
\end{equation}
which extends Eq. (\ref{wdtintdifec0}) to finite densities. The
dashed line in Fig. 1 corresponds to this expression with the value $D^*$=0.099. The comparison with the simulation data indicates that Eq. (\ref{wdtstar1}) provides an accurate representation of $W(t)$ at short and at long times compared with $\tau_0$. In this comparison, the parameter $D^*$ was employed as a mere fitting parameter. The idea, however, is to build a first-principles theory that {\it predicts} the value of $D^*$ and the behavior of $W(t)$ in the intermediate-time regime. The results of the present paper will eventually constitute an essential ingredient in the construction of the \emph{atomic }version of the self-consistent GLE theory. The most immediate task, however, is to provide a more formal understanding of the physical meaning and the fundamental
nature of the kinetic friction represented by $\zeta^0$, and
in general, of the Langevin equation for atomic liquids in Eq.
(\ref{langevin0}). This task is addressed in the following two sections.

\section{Kinetic friction on an atomic tracer particle.}\label{sectionIII}

The fact that the Langevin equation in Eq. (\ref{langevin0}) adequately describes the results for the MSD of Newtonian systems strongly suggests a more profound equivalence between the dynamics of Newtonian and Brownian liquids, at least regarding tracer diffusion phenomena. In both cases the relaxation time $\tau_S \equiv [M/\zeta^0]$ of the velocity, due to the friction force $-\zeta^0{\bf v}(t)$, defines the crossover from ballistic ($t\ll\tau_S$) to diffusive ($t\gg\tau_S$) motion. The fundamental difference lies in the physical origin of the friction force  $-\zeta^0{\bf v}(t)$: in a Brownian liquid, this friction is assumed to be caused by an external material agent, namely, the supporting solvent, which also acts as a heat reservoir. In a Newtonian liquid, in contrast, the (`Doppler') friction force  $-\zeta^0{\bf v}(t)$ is not caused by any external material agent but by the molecular collisions responsible to impose the equipartition of energy among  the kinetic energy degrees of freedom of the system.

In other words, the underlying physical origin of the kinetic friction effects is that for times $t$ much longer than $\tau_0$, in which each particle has undergone many molecular collisions, the kinetic energy has indeed been distributed according to the most probable distribution. This then means that partial thermal equilibrium, consisting of this equipartition of the kinetic energy, is
achieved within the time scale represented by $\tau_0$. Such
partial thermalization must involve the transport of heat
through molecular collisions. The instantaneous
fluctuations of this local temperature, however, lead to
the random force $\textbf{f}^0(t)$ that cause the instantaneous fluctuations in the velocity of the tracer particle. The same molecular mechanisms are
also responsible for the emergence of the systematic friction force
$-\zeta^0\overline{{\bf v}(t)}$ on the tracer particle whenever this particle
has a non-zero mean velocity $\overline{{\bf v}(t)}$.

Let us now develop a more microscopic and precise discussion of these physical effects. For this, let us now consider $(N+1)$ particles in a volume $V$, the first of them (the tracer particle) having momentum $\mathbf{p}(t)=M\mathbf{v}(t)$ and the rest $N$ particles having momentum $\mathbf{p}_i (t)=m\mathbf{v}_i(t)$, with $1\le i\le N$. In the absence of external forces, the total momentum $\mathbf{P}_{total} \equiv \mathbf{p}(t) + \sum_{i=1}^N  \mathbf{p}_i (t)$ is conserved,

\begin{equation}
\frac{d}{dt}[\mathbf{p}(t) + \sum_{i=1}^N  \mathbf{p}_i (t)] =0,
\end{equation}
or

\begin{equation}
\frac{d}{dt}[\mathbf{p}(t) + \int _{V'} \mathbf{j}(\mathbf{r},t) \ d^3r ] =0,
\end{equation}
with

\begin{equation}
\mathbf{j}(\mathbf{r},t) \equiv  \sum_{i=1}^N  \mathbf{p}_i (t) \delta (\mathbf{r}-\mathbf{r}_i(t)),
\label{current}
\end{equation}
and with $V'\equiv V-V_T$ being the total confining volume $V$ minus the excluded volume $V_T$ around the center of the tracer particle. Thus, $V'$ is bounded by an outer surface $\Sigma$ of the confining walls, which remain fixed in space, and by the surface $\sigma$ around $V'$, which  follows the motion of this particle, and hence, is not fixed in space.  Clearly, then, the momentum conservation equation can also be written as
\begin{equation}
\frac{d\mathbf{p}(t)}{dt} = - \int _{V'}  \left( \frac{\partial \mathbf{j}(\mathbf{r},t)}{\partial t}\right) \ d^3r.
\label{cont0}
\end{equation}

On the other hand, taking the time derivative of Eq. (\ref{current}), one can write the continuity equation for the momentum density $ \mathbf{j}(\mathbf{r},t)$ as
\begin{equation}
 \left( \frac{\partial \mathbf{j}(\mathbf{r},t)}{\partial t}\right) = \nabla \cdot \stackrel {\leftrightarrow}{\Pi}_K  (\mathbf{r},t)+ \nabla \cdot \stackrel {\leftrightarrow}{\Pi}_U  (\mathbf{r},t) ,
 \label{cont1}
\end{equation}
where $\stackrel {\leftrightarrow}{\Pi}_K  (\mathbf{r},t)$ and $\stackrel {\leftrightarrow}{\Pi}_U  (\mathbf{r},t)$ are the kinetic and configurational components of the stress tensor $\stackrel {\leftrightarrow}{\Pi} (\mathbf{r},t)$, i.e.,
\begin{equation}
\stackrel {\leftrightarrow}{\Pi} (\mathbf{r},t)=\stackrel {\leftrightarrow}{\Pi}_K (\mathbf{r},t) +\stackrel {\leftrightarrow}{\Pi}_U (\mathbf{r},t).
\end{equation}
The kinetic part $\stackrel {\leftrightarrow}{\Pi}_K  (\mathbf{r},t)$, which describes  the  change  of  momentum  due  to
particles  crossing  the  boundaries of $V'$, is a  second rank tensor with components \cite{schofieldhenderson}
\begin{equation}
 \left[ \stackrel {\leftrightarrow}{\Pi}_K  (\mathbf{r},t)\right]^{\alpha \beta} \equiv -\sum _{i=1}^N \frac{p_i^\alpha(t) p_i^\beta(t)}{m}\delta(\mathbf{r}-\mathbf{r}_i(t)). \label{pikin}
\end{equation}
The configurational part of $\stackrel {\leftrightarrow}{\Pi} (\mathbf{r},t)$, on the other hand, is defined by the condition
\begin{equation}
\nabla \cdot \stackrel {\leftrightarrow}{\Pi}_U  (\mathbf{r},t) \equiv \sum_{i=1}^N  \dot{{\mathbf{p}}}_i (t) \delta (\mathbf{r}-\mathbf{r}_i(t)),
\end{equation}
so that the continuity equation for $ \mathbf{j}(\mathbf{r},t)$, Eq. (\ref{cont1}), can actually be written most simply as
\begin{equation}
 \left( \frac{\partial \mathbf{j}(\mathbf{r},t)}{\partial t}\right) = \nabla \cdot \stackrel {\leftrightarrow}{\Pi}_K  (\mathbf{r},t)+ \sum_{i=1}^N  \dot{{\mathbf{p}}}_i (t) \delta (\mathbf{r}-\mathbf{r}_i(t)).
 \label{cont10}
\end{equation}

Thus, substituting this equation in Eq. (\ref{cont0}), we have
\begin{equation}
\frac{d\mathbf{p}(t)}{dt} = -\int_{V'} \left( \nabla  \cdot \stackrel {\leftrightarrow}{\Pi}_K (\mathbf{r},t)\right) \ d^3r - \int _{V'}  \left(   \sum_{i=1}^N  \dot{{\mathbf{p}}}_i (t) \delta (\mathbf{r}-\mathbf{r}_i(t))  \right) \ d^3r.
 \label{cont2}
\end{equation}

The second term on the right side of Eq. (\ref{cont2}) is just $-\sum_{i=1}^N  \dot{{\mathbf{p}}}_i (t)= -\sum_{i=1}^N {\mathbf{F}}_i (t)= -\sum_{i=1}^N [\sum_{j\ne i}^N{\mathbf{F}}_{ij} (t)+ {\mathbf{F}}_{iT} (t)]=-\sum_{i=1}^N {\mathbf{F}}_{iT} (t)=\sum_{i=1}^N {\mathbf{F}}_{Ti} (t)$, where ${\mathbf{F}}_{ij} (t)$ is the force exerted on particle $i$ by particle $j$ and ${\mathbf{F}}_{Ti} (t)$ is the force exerted \emph{by} particle $i$ on the tracer particle, and where we have used the fact that ${\mathbf{F}}_{ij} (t)=-{\mathbf{F}}_{ji} (t)$. We may then write this term as $\sum_{i=1}^N{\mathbf{F}}_{Ti} (t)=\sum_{i=1}^N\nabla_i u(\mid \textbf{r}_i-\textbf{r}_T\mid)=\int [\nabla_{\mathbf{r}} u(\mid \textbf{r}-\textbf{r}_T(t)\mid)]\sum_{i=1}^N \delta (\mathbf{r}-\mathbf{r}_i(t)) \ d^3r$. By shifting the origin of the coordinate system to the center of the tracer particle (including the variable of integration $\textbf{r}$), it is not difficult to see that Eq. (\ref{cont2}) can be rewritten as
\begin{equation}
\frac{d\mathbf{p}(t)}{dt} = -\int_{V'} \left( \nabla  \cdot \stackrel {\leftrightarrow}{\Pi}_K (\mathbf{r},t)\right) \ d^3r+ \int _{V'}  \left [  \nabla u(r)  \right] n^*(\textbf{r},t)\ d^3r,
 \label{cont3}
\end{equation}
with
\begin{equation}
n^*(\mathbf{r},t) \equiv  \sum_{i=1}^N   \delta (\mathbf{r}-\mathbf{r}_i(t))
\label{nstar}
\end{equation}
being the local particle density around the tracer particle described, however, from a reference frame \emph{whose origin moves} together with the center of this particle.

Eq. (\ref{cont3}) shows that there is a very simple and exact coupling between the force on the tracer particle and two collective variables  of the surrounding fluid, namely, the kinetic component $\stackrel {\leftrightarrow}{\Pi}_K (\mathbf{r},t)$ of its stress tensor (whose trace is a measure of the instantaneous local kinetic energy per particle, i.e., of the local instantaneous temperature) and the local number density $n^*(\textbf{r},t)$. Thus, the interatomic forces affect the motion of a tracer particle through two physically distinct channels. The first of them is a kinetic constraint imposed by momentum conservation, and the second is the ordinary configurational effect of interatomic forces. Each of these two variables contribute additively and linearly to the total force on the tracer particle.

Taking the equilibrium average (indicated by an overbar) of Eq. (\ref{cont3}), we have that
\begin{equation}
\frac{d\overline{\mathbf{p}}}{dt} = -\int_{V'} \left( \nabla  \cdot\overline{ \stackrel {\leftrightarrow}{\Pi}_K (\mathbf{r})}\right) \ d^3r+ \int _{V'}  \left [  \nabla u(r)  \right]\overline{  n^*(\textbf{r})}\ d^3r,
 \label{meancont3}
\end{equation}
where the mean value of $\stackrel {\leftrightarrow}{\Pi}_K (\mathbf{r},t)$ can be obtained averaging Eq. (\ref{pikin}), with the result
\begin{equation}
\overline{ \stackrel {\leftrightarrow}{\Pi}_K (\mathbf{r})}=-(k_BT)\stackrel {\leftrightarrow}{\textbf{I}}\overline{  n^*(\textbf{r})},
 \label{meanpikin}
\end{equation}
with $\stackrel{\leftrightarrow }{{\bf I}}$ being the 3$\times$3 cartesian unit tensor and  with $\overline{  n^*(\textbf{r})}$ given by
\begin{equation}
\overline{  n^*(\textbf{r})}= \overline{n} g(r),
 \label{meanlocaln}
\end{equation}
where $g(r)$ is the bulk radial distribution function of the system. From these results it is not difficult to realize that for symmetry reasons each of the two mean forces on the right side of Eq. (\ref{meancont3}) vanish independently, so that the tracer particle experiences a vanishing total mean force, and  $(d\overline{\mathbf{p}}/dt)=\textbf{0}$.

We may now write the state variables $\stackrel {\leftrightarrow}{\Pi}_K (\mathbf{r})$ and $ n^*(\textbf{r})$ as the sum of their equilibrium mean value plus the corresponding fluctuations, namely, as
\begin{equation}
\stackrel {\leftrightarrow}{\Pi}_K (\mathbf{r})=\overline{ \stackrel {\leftrightarrow}{\Pi}_K (\mathbf{r})}+\delta\stackrel {\leftrightarrow}{\Pi}_K (\mathbf{r})
 \label{dpikin}
\end{equation}
and
\begin{equation}
 n^*(\textbf{r})= \overline{  n^*(\textbf{r})}+ \delta n^*(\textbf{r}).
 \label{dmeanlocal}
\end{equation}
This allows us to rewrite Eq. (\ref{cont3}) as an exact relationship between these thermal fluctuations and the instantaneous momentum of the particle, namely
\begin{equation}
\frac{d\mathbf{p}(t)}{dt} = -\int_{V'} \left( \nabla  \cdot \delta \stackrel {\leftrightarrow}{\Pi}_K (\mathbf{r},t)\right) \ d^3r+ \int _{V'}  \left [  \nabla u(r)  \right]\delta n^*(\textbf{r},t)\ d^3r.
 \label{contfluct}
\end{equation}
This exact equation will now be taken as the starting point for a formal statistical mechanical derivation of the ordinary Langevin equation of a the tracer particle in our atomic liquid.

For this, let us  first recall that the basis of the GLE formalism are the general mathematical conditions stated by the theorem of stationarity \cite{delrio}. This theorem states that the equation describing the dynamics of the thermal fluctuations $\delta a_{i}(t)\ (\equiv a_{i}(t)-a^{eq}_{i})$ of the instantaneous value of the macroscopic variables $ a_{i}(t)$ ($i=1,2,...,\nu$) around its equilibrium value $a^{eq}_{i}$ must have the structure of the most general linear stochastic equation with additive noise for the vector $\delta \mathbf{a}(t)=\left[\delta a_{1}(t),\delta a_{2}(t),...,\delta a_{\nu }(t)\right]^\dagger $, namely,
\begin{equation}
\frac{d\delta \mathbf{a}(t)}{dt}=-\omega \chi ^{-1}\delta \mathbf{a}(t)-\int%
\limits_{0}^{t}L(t-t^{\prime })\chi ^{-1}\delta \mathbf{a}(t^{\prime })dt^{\prime }+%
\mathbf{f}(t).
\label{gle0}
\end{equation}
In this equation $\chi $ is the matrix of static correlations,  $\chi
_{ij}\equiv \left\langle \delta  a_{i}(0) \delta  a_{j}^{\ast }(0)\right\rangle $, $\omega $
is an anti-Hermitian matrix ($\omega _{ij}=-\omega _{ji}^{\ast }$), and the matrix $L(t)$
is determined by the fluctuation-dissipation relation $L_{ij}(t)=\left\langle
f_{i}(t)f_{j}^*(0)\right\rangle $, where $f_{i}(t)$ is the $i$th component of the vector of random forces $\mathbf{f}(t)$. Besides the selection rules imposed by these symmetry properties of the matrices  $\chi $, $\omega $, and $L(t)$, other selection rules are imposed by other symmetry conditions. For example \cite{delrio}, if the variables $ a_{i}(t)$ have a definite parity upon time reversal, $ a_{i}(-t)=\lambda_i a_{i}(t)$ with  $ \lambda_i =$ 1 or -1, then
$\omega _{ij}=-\lambda_i\lambda_j\omega _{ij}$ and $L _{ij}(t)=\lambda_i\lambda_j L _{ij}(t)$.

Let us now apply this mathematical infrastructure to the physical context involving the exact momentum conservation equation, Eq. (\ref{cont3}), and let us define the vector $\delta \mathbf{a}(t)$, partitioned as
\begin{equation}
\delta \mathbf{a}(t)=\left[\mathbf{p}(t),\delta \Pi (t), \delta n^*(t)
\right]^\dagger,
\label{vectora}
\end{equation}
in terms of the sub-vectors $\mathbf{p}(t),\delta \Pi(t)$, and $\delta n^*(t)$, defined by their components
\begin{equation}
\mathbf{p}(t)=\left(p_x(t), p_y(t), p_z(t)\right),
\label{vectorp}
\end{equation}
\begin{equation}
\left[\delta \Pi (t)\right]^{\alpha \beta}(\textbf{r}) = \delta \Pi_K ^{\alpha \beta}(\textbf{r},t) \ \ \  \ (\textrm{with}\ \ \ \alpha, \beta =x,y,z,\ \ \ \ \textrm{and} \ \ \ \textbf{r}\in V)
\label{vectorpikin}
\end{equation}
and
\begin{equation}
\left[\delta n^*(t)\right](\textbf{r}) = \delta n^*(\textbf{r},t) \ \ \  \ (\textrm{with} \ \ \ \textbf{r}\in V).
\label{vectorpikin}
\end{equation}

With this definition of the vector $\delta \mathbf{a}(t)$ one can calculate the static correlation matrix $\chi
\equiv \left\langle \delta \textbf{a}(0) \delta \textbf{a}^{\dagger }(0)\right\rangle $ using the microscopic definitions of $\delta n^*(t)$ in Eqs. (\ref{nstar}), (\ref{meanlocaln}),  and (\ref{dmeanlocal}) and of $\delta \Pi_K ^{\alpha \beta}(\textbf{r},t)$ in  Eqs. (\ref{pikin}), (\ref{meanpikin}),  and (\ref{dpikin}). The result for $\chi$ can be written as the following partitioned matrix
\begin{equation}
{\bf \chi }=\left[
\begin{array}{cccc}
\chi _{pp} & 0 & 0  \\
0 & \chi _{\pi\pi} & \chi _{\pi n}  \\
0 & \chi _{n\pi} & \chi _{nn}
\end{array}
\right],
\end{equation}
whose sub-matrices have elements defined as $[\chi _{pp}]^{\alpha \beta}\equiv \langle p^\alpha p^\beta \rangle $, $[\chi _{\pi\pi}]^{\alpha \beta,\mu \nu}(\textbf{r},\textbf{r}')\equiv \langle \delta \Pi_K ^{\alpha \beta}(\textbf{r})\delta \Pi_K ^{\mu \nu}(\textbf{r}')\rangle $, $[\chi _{\pi n}]^{\alpha \beta}(\textbf{r},\textbf{r}')\equiv \langle \delta \Pi_K ^{\alpha \beta}(\textbf{r})\delta n^*(\textbf{r}')\rangle $,
$[\chi _{n\pi}]^{\mu \nu}(\textbf{r},\textbf{r}')\equiv \langle \delta n^*(\textbf{r})\delta \Pi_K ^{\mu \nu}(\textbf{r}')\rangle $, and $[\chi _{n n}](\textbf{r},\textbf{r}')\equiv \langle \delta n^*(\textbf{r})\delta n^*(\textbf{r}')\rangle $, given, respectively, by
\begin{equation}
[\chi _{pp}]^{\alpha \beta}=(Mk_{B}T) \delta _{\alpha \beta},  \label{chipp}
\end{equation}
\begin{equation}
[\chi _{\pi\pi}]^{\alpha \beta,\mu \nu}(\textbf{r},\textbf{r}')=(k_{B}T)^2\{[ \delta _{\alpha \beta}\delta _{\mu \nu} +\delta _{\alpha \mu}\delta _{\beta \nu} +\delta _{\alpha \nu}\delta _{\mu \beta}]\chi_s(\textbf{r},\textbf{r}') + \delta _{\alpha \beta}\delta _{\mu \nu}\chi_d(\textbf{r},\textbf{r}')\},  \label{chipipi}
\end{equation}
\begin{equation}
[\chi _{\pi n}]^{\alpha \beta}(\textbf{r},\textbf{r}')=[\chi _{n \pi }]^{\alpha \beta}(\textbf{r},\textbf{r}')=-(k_{B}T) \delta _{\alpha \beta}\chi_d(\textbf{r},\textbf{r}'), \label{chipin}
\end{equation}
and
\begin{equation}
[\chi _{nn}](\textbf{r},\textbf{r}')=\chi_s(\textbf{r},\textbf{r}') + \chi_d(\textbf{r},\textbf{r}'), \label{chinn}
\end{equation}
where the self and the distinct parts of $\chi _{nn}$ are defined as
\begin{equation}
\chi_s(\textbf{r},\textbf{r}') = \overline{  n^*(\textbf{r})} \delta(\textbf{r}-\textbf{r}') \label{chis}
\end{equation}
and
\begin{equation}
\chi_d(\textbf{r},\textbf{r}') = \overline{  n^*(\textbf{r})  n^*(\textbf{r})}-\overline{  n^*(\textbf{r})}\ \overline{  n^*(\textbf{r})}   \label{chid}
\end{equation}

We then write up the generalized Langevin equation for our vector $\delta {\bf a}(t)$ in the format
of Eq. (\ref{gle0}). For this, we first
notice that all the variables, except $\textbf{p}(t)$, are
even functions under time-reversal. According to Onsager's reciprocity
relations, and the general anti-hermiticity of $\omega $ and hermiticity of $
L(t)$  \cite{delrio}, we have that the only possibly non-zero submatrices of $\omega $ and $L(t)$ are

\begin{equation}
\omega {\bf =}\left[
\begin{array}{cccc}
0 & \omega _{p\pi} & \omega _{pn} \\
-\omega _{p \pi }^{\dagger} & 0 & 0 \\
-\omega _{p n}^{\dagger } & 0 & 0
\end{array}
\right]
\end{equation}
and
\begin{equation}
L(t)=\left[
\begin{array}{cccc}
L_{pp}(t) & 0 & 0 \\
0 & L_{\pi \pi}(t) & L_{\pi n}(t) \\
0 & L_{n \pi}^{\dagger}(t) & L_{nn}(t)
\end{array}
\right].
\end{equation}

The determination of some of the non-zero elements of $\omega $ and $L(t)$ is rather straightforward. Thus the previous selection rules, along with the general format imposed by the GLE equation (\ref{gle0}), allows us to write the time-evolution equation for the sub-vector $\textbf{p}(t)$ as
\begin{eqnarray}
\frac{d\mathbf{p}(t)}{dt} &=&-\left[\omega_{p\pi}(\chi^{-1})_{\pi\pi} +  \omega_{p n}(\chi^{-1})_{n\pi}  \right]\delta\Pi (t)-\left[\omega_{p\pi}(\chi^{-1})_{\pi n} +  \omega_{p n}(\chi^{-1})_{n n}  \right]\delta n^* (t)
\nonumber \\
&&-\int_0^t dt' L_{pp}(t-t')(\chi^{-1})_{pp}\mathbf{p}(t') + f_p(t) \label{dpdtgral}
\end{eqnarray}
By comparing with the exact momentum conservation equation in Eq. (\ref{dpdtgral}), we immediately conclude that
$L_{pp}(t)=f_p(t)=0$, and that the remaining terms correspond, respectively, to the kinetic and configurational forces on the right side of this equation. In addition, for simplicity we approximate $\overline{  n^*(\textbf{r})  n^*(\textbf{r})}\approx \overline{  n^*(\textbf{r})}\ \overline{  n^*(\textbf{r})}$ in Eqs. (\ref{chipin})  and (\ref{chid}), so as to neglect at this point the static cross-correlation $\chi_{n\pi}$ and $\chi_{\pi n}$,
\begin{equation}
[\chi _{\pi n}]^{\alpha \beta}(\textbf{r},\textbf{r}')=[\chi _{n \pi }]^{\alpha \beta}(\textbf{r},\textbf{r}')\approx 0,
\end{equation}
so that the previous equation is rewritten as
\begin{equation}
\frac{d\mathbf{p}(t)}{dt} = -\omega_{p\pi}(\chi^{-1})_{\pi\pi} \delta\Pi (t)- \omega_{p n}(\chi^{-1})_{n n}  \delta n^* (t).
\label{dpdtgralp}
\end{equation}
Comparing this equation with Eq. (\ref{dpdtgral}) one can determine the sub-matrices $\omega_{p\pi}$ and $\omega_{pn}$ and hence, also the sub-matrices $\omega_{\pi p}\ (= - \omega_{p\pi}^\dagger)$ and $\omega_{np}\ (= - \omega_{pn}^\dagger)$.

In a  similar manner, from the exact format imposed by the GLE, and using the previous selection rules (as well as the approximation $\chi_{n\pi}\approx 0$), one can also write the time-evolution equations for the other two variables, $\delta\Pi (t)$ and $\delta n^*(t)$, as
\begin{eqnarray}
\frac{d\delta\Pi(t)}{dt} &=&-\omega_{\pi p}\chi^{-1}_{pp} \mathbf{p}(t) -\int_0^t dt' L_{\pi \pi}(t-t')(\chi^{-1})_{\pi \pi}\delta\Pi(t')
\nonumber \\
&&-\int_0^t dt' L_{\pi n}(t-t')(\chi^{-1})_{nn}\delta n^*(t')  + f_\pi(t) \label{dpidtgralp}
\end{eqnarray}
and
\begin{eqnarray}
\frac{d \delta n^*(t)}{dt} &=&-\omega_{n p}\chi^{-1}_{pp} \mathbf{p}(t) -\int_0^t dt' L_{n \pi}(t-t')(\chi^{-1})_{\pi \pi}\delta\Pi(t')
\nonumber \\
&&-\int_0^t dt' L_{n n}(t-t')(\chi^{-1})_{nn}\delta n^*(t')  + f_n(t) \label{dndtgralp}
\end{eqnarray}

Eqs. (\ref{dpdtgralp})-(\ref{dndtgralp}) provide a non-contracted description of the thermal fluctuations in an atomic liquid, which involves the tracer particle's momentum $\mathbf{p}(t)$ as one of the variables. Contracting this description to the state subspace spanned by $\mathbf{p}(t)$ itself will finally lead to the complete generalized Langevin equation for a tracer particle in such atomic liquid. The result, however, is rather involved, but the essence may be best appreciated if we introduce an additional simplification, which consists of neglecting the dissipative coupling between the variables  $\delta\Pi (t)$ and $\delta n^*(t)$, i.e., by setting $L_{\pi n}(t)=L_{n\pi}(t)=0.$ Under these circumstances, the solution of Eq. (\ref{dpidtgralp}) can be written as
\begin{eqnarray}
\delta\Pi(t) &=&G_\pi(t)\delta\Pi(0)-\int_0^t dt' G_\pi(t-t')\omega_{\pi p}\chi^{-1}_{pp} \mathbf{p}(t')
\nonumber \\
&&+\int_0^t dt' G_\pi(t-t')f_\pi(t') \label{funcgreenpi}
\end{eqnarray}
where the Green's function $G_\pi(t)$ is the solution of
\begin{equation}
\frac{dG_\pi(t)}{dt} = -\int_0^t dt' L_{\pi \pi}(t-t')(\chi^{-1})_{\pi \pi}G_\pi(t'),
\label{eqgreen}
\end{equation}
with initial condition $G_\pi(t=0)=I$, i.e., its Laplace transform (LT) $\hat{G}_\pi(z)$  will be given, in terms of the LT of $L_{\pi \pi}(t)$, by
\begin{equation}
\hat{G}_\pi(z) = \left[zI+\hat{L}_{\pi \pi}(z)(\chi^{-1})_{\pi \pi}\right]^{-1}.
\label{ltgreen}
\end{equation}

Substituting the expression above for $\delta\Pi(t)$ in Eq. (\ref{dpdtgralp}) we arrive at the following Langevin equation
\begin{equation}
\frac{d\mathbf{p}(t)}{dt} = -\frac{1}{M}\int_0^t dt'  \stackrel {\leftrightarrow}{\zeta}   _K (t-t')\cdot \mathbf{p}(t') +  \mathbf{f}_K(t)
- \omega_{p n}(\chi^{-1})_{n n}  \delta n^* (t),
\label{dpdtgralpp}
\end{equation}
where we have defined the time-dependent kinetic friction coefficient $ \zeta _K (t)$ as
\begin{equation}
\frac{  \stackrel {\leftrightarrow}{\zeta} _K (t)}{M}\equiv -\omega_{p\pi}(\chi^{-1})_{\pi\pi}  G_\pi(t)\omega_{\pi p}\chi^{-1}_{pp}  \label{deltazetak}
\end{equation}
and the kinetic random force $\mathbf{f}_K(t)$ as
\begin{equation}
\mathbf{f}_K(t)\equiv -\omega_{p\pi}(\chi^{-1})_{\pi\pi} \left[ G_\pi(t)\delta\Pi(0)
+\int_0^t dt' G_\pi(t-t')f_\pi(t')\right]. \label{funcgreenf}
\end{equation}
According to the theorem of contractions \cite{delrio}, $\mathbf{f}_K(t)$ and $ \stackrel {\leftrightarrow}{\zeta}  _K (t)$ must satisfy the  fluctuation-dissipation relationship $\langle\mathbf{f}_K^{\alpha}(t)\mathbf{f}_K^{\beta}(t')\rangle= k_BT \zeta _K ^{\alpha\beta}(t-t')$.

For future reference, let us notice that Eq. (\ref{ltgreen}) allows us to write the FT of the kinetic friction function $ \stackrel {\leftrightarrow}{\zeta} _K (t)$ in Eq. (\ref{deltazetak}) directly in terms of the FT  of the memory function $L_{\pi \pi}(t)$ as
\begin{equation}
\frac{  \stackrel {\leftrightarrow}{\zeta} _K (z)}{M}\equiv -\omega_{p\pi}(\chi^{-1})_{\pi\pi} \left[zI+\hat{L}_{\pi \pi}(z)(\chi^{-1})_{\pi \pi}\right]^{-1} \omega_{\pi p}\chi^{-1}_{pp}.  \label{deltazetakp}
\end{equation}
Clearly, the exact determination of the memory function $L_{\pi \pi}(t)$, and hence, of the kinetic friction coefficient  $  \stackrel {\leftrightarrow}{\zeta}  _K (t)$, is perhaps impossible, but some properties can be drawn from the expressions just derived, at least in certain limits and within well-defined approximations. For example, if  $  \stackrel {\leftrightarrow}{\zeta}  _K (t)$ relaxes to zero within a finite relaxation time, then for times $t$ much longer than such relaxation time we can approximate  $  \stackrel {\leftrightarrow}{\zeta}  _K (t)$ by its Markov limit,
\begin{equation}
 \stackrel {\leftrightarrow}{\zeta}  _K (t)\approx  2\delta(t) \stackrel {\leftrightarrow}{\zeta ^0} ,  \label{markov}
\end{equation}
where
\begin{equation}
 \stackrel {\leftrightarrow}{\zeta^0} = \int_0^\infty dt   \stackrel {\leftrightarrow}{\zeta}  _K (t).  \label{zetamarkov}
\end{equation}
In addition, due to the radial symmetry of the interparticle interactions,  $\stackrel {\leftrightarrow}{\zeta^0}$ must be isotropic (i.e., diagonal), so that
\begin{equation}
 \stackrel {\leftrightarrow}{\zeta^0} = \zeta^0\stackrel {\leftrightarrow}{I},  \label{zetamarkovdiag}
\end{equation} with $ \zeta^0$ given by
\begin{equation}
\zeta^0 =    -M\left[\omega_{p\pi} \hat{L}_{\pi \pi}^{-1}(z=0) \omega_{\pi p}\chi^{-1}_{pp}\right]^{xx}.
\label{zetamarkov}
\end{equation}

Thus, we conclude that in the Markov limit Eq. (\ref{dpdtgralpp}) can be written as
\begin{equation}
M\frac{d\mathbf{v}(t)}{dt} = -\zeta^0 {\bf v}(t)+{\bf f}^0 (t) - \omega_{p n}(\chi^{-1})_{n n}  \delta n^* (t).
\label{dpdtgralppp}
\end{equation}
In the following section we discuss the additional contraction process that leads to the elimination of the variable $\delta n^* (t)$ from the description consisting of this equation and of Eq. (\ref{dndtgralp}) above. Before that, however, let us mention that
Eq. (\ref{deltazetakp}), and in particular its Markov limit in Eq. (\ref{zetamarkov}), involves the  LT of the memory function $\hat{L}_{\pi \pi}^{-1}(z)$ as the only unknown quantity, which the GLE formalism is unable to determine. Although one could introduce additional approximations to determine this memory function, this is not the main objective of the present paper; instead, the derivation above was only meant to provide a more formal explanation of the origin of the kinetic friction and random forces, introduced and discussed in more efficient and intuitive terms in the previous section. After all, such arguments did provide a simple and accurate zeroth-order approximate  determination of the kinetic friction coefficient $ \zeta^0$, namely, the use of the kinetic-theory value of the self-diffusion coefficient, Eq. (\ref{dkinetictheory}), in Einstein's relation, Eq. (\ref{einstein}).

\section{{Configurational friction on an atomic tracer particle.}}\label{sectionIV}

One important contribution of the previous section was to make a point that in an atomic liquid the force on a tracer particle couples linearly with the kinetic component $\stackrel {\leftrightarrow}{\Pi}_K (\mathbf{r},t)$ of the stress tensor and with the local number density $n^*(\textbf{r},t)$ of the surrounding fluid, as indicated by Eq. (\ref{cont3}). As a consequence, the interatomic forces affect the motion of a tracer particle through two physically distinct channels, namely, the kinetic constraint imposed by momentum conservation and the ordinary configurational effect of interatomic forces. Another important conclusion was to notice that the former is the origin of the kinetic friction, finally formatted in Eq. (\ref{dpdtgralppp}) as a dissipative friction term $-\zeta^0 {\bf v}(t)$ plus the corresponding random force ${\bf f}^0 (t)$. Let us now discuss the effects of the coupling with  $\delta n^*(\textbf{r},t)$.

For this, let us resume the formal process of contraction of the description initiated in the previous section. We thus recall that after projecting out the variable $\delta \stackrel{\leftrightarrow}{\Pi}_K (\mathbf{r},t)$, the time-evolution equations for the remaining fluctuating variables are eqs. (\ref{dpdtgralpp}) and (\ref{dndtgralp}). For clarity, we rewrite here these equations as
\begin{equation}
\frac{d\mathbf{p}(t)}{dt} = -\frac{\zeta^0}{M} {\bf p}(t)+{\bf f}^0 (t)
- \omega_{p n}(\chi^{-1})_{n n}  \delta n^* (t),
\label{dpdtgralpppp}
\end{equation}
and
\begin{eqnarray}
\frac{d \delta n^*(t)}{dt} = -\omega_{n p}\chi^{-1}_{pp} \mathbf{p}(t) -\int_0^t dt' L_{n n}(t-t')(\chi^{-1})_{nn}\delta n^*(t')  + f_n(t), \label{dndtgralpp}
\end{eqnarray}
where the kinetic friction term of eq. (\ref{dpdtgralpp}) has been written in its markov limit (as in Eq. (\ref{dpdtgralppp})), and where the term of  eq. (\ref{dndtgralp}) involving $L_{n\pi}(t)$ has been neglected, as discussed immediately above Eq. (\ref{funcgreenpi}). We now formally project out the variable $\delta n^*(t)$  by solving Eq. (\ref{dndtgralpp}) for this variable, and substituting the resulting solution in the third term of the right side of Eq. (\ref{dpdtgralpppp}).

This contraction process results in the following generalized Langevin equation for the velocity ${\bf v}(t)\ (={\bf p}(t)/M)$ of a tracer particle in the atomic liquid,
\begin{equation}
M{\frac{d{\bf v}(t)}{dt}}= -\zeta^0 {\bf v}(t)+{\bf f}
^0(t)- \int_0^t dt' \stackrel{\leftrightarrow }{\Delta
\zeta(t}-t') \cdot {\bf v}(t')+ {\bf F} (t), \label{eq6}
\end{equation}
where the\emph{ configurational} time-dependent friction function $\stackrel{\leftrightarrow }{\Delta
\zeta(t})$ is given by
\begin{equation}
\frac{  \Delta \stackrel {\leftrightarrow}{\zeta} (t)}{M}\equiv -\omega_{pn}(\chi^{-1})_{nn}  G(t)\omega_{n p}\chi^{-1}_{pp}, \label{deltazetaconf}
\end{equation}
and the new (configurational) random force $\mathbf{F}(t)$ as
\begin{equation}
\mathbf{F}(t)\equiv -\omega_{pn}(\chi^{-1})_{nn} \left[ G(t)\delta n^*(0)
+\int_0^t dt' G(t-t')f_n(t')\right]. \label{funcgreenconf}
\end{equation}
In these equations, the Green's function $G(t)$ is the solution of
\begin{eqnarray}
\frac{d G(t)}{dt} =  -\int_0^t dt' L_{n n}(t-t')(\chi^{-1})_{nn}G(t') \label{greenfunctconf}
\end{eqnarray}
with initial condition $G(0)=I$. According to the contraction theorem \cite{delrio}, $\mathbf{F}(t)$ and $\Delta \stackrel {\leftrightarrow}{\zeta} (t)$ must satisfy the  fluctuation-dissipation relationship $\langle F^{\alpha}(t)F^{\beta}(t')\rangle= k_BT \Delta  \zeta ^{\alpha\beta}(t-t')$.

Although the previous statements are physically accurate and well-defined, it is also useful to rephrase this abstract derivation in more concrete and intuitive terms. For this we rewrite Eq. (\ref{dpdtgralpppp})  as
\begin{equation}
M\frac{d\mathbf{v}(t)}{dt} = -\zeta^0 {\bf v}(t)+{\bf f}^0 (t)+ \int d^3{\bf r} [\nabla u(r)]\delta n^*(\textbf{r},t), \label{eq3}
\end{equation}
to recover the original notation in Eq. (\ref{contfluct}) for the configurational force term. The comparison with Eq. (\ref{dpdtgralpppp}) then implies that $[- \omega_{p n}(\chi^{-1})_{n n} ]^{\alpha}(\textbf{r}) = [\nabla^\alpha u(r)]$ and, convoluting this equation with $[\chi _{n n}](\textbf{r},\textbf{r}')\equiv \langle \delta n^*(\textbf{r})\delta n^*(\textbf{r}')\rangle $ determines that $\omega_{p n}$ is given by
\begin{eqnarray}
[ \omega_{p n} ]^{\alpha}(\textbf{r}) && = - \int d^3r' [\nabla^{'\alpha} u(r')]\chi _{n n}(\textbf{r}',\textbf{r}) \nonumber \\
&& = k_BT [ \nabla^{\alpha}n^{eq}(\textbf{r})],
\end{eqnarray}
where the second equality is a direct consequence of the exact equilibrium condition referred to as the Wertheim-Lovett's relation, namely \cite{evans},
\begin{eqnarray}
[ \nabla^{\alpha}n^{eq}(\textbf{r})]= - \beta \int d^3r' \chi _{n n}(\textbf{r}',\textbf{r})[\nabla^{'\alpha} u(r')].
\end{eqnarray}
In the previous equations, the equilibrium mean value $\overline{n}(\textbf{r})$ has been denoted by  $n^{eq}(\textbf{r})$.

Now, since  $\omega_{n p}=- [\omega_{p n}]^\dagger$, we find that $[ \omega_{np} ]^{\alpha}(\textbf{r}) =- k_BT [ \nabla^{\alpha}n^{eq}(\textbf{r})]$. Using this result, along with the value $[\chi _{pp}]^{\alpha \beta}=(Mk_{B}T) \delta _{\alpha \beta}$ (Eq. (\ref{chipp})), we can write Eq. (\ref{dndtgralpp}) more concretely as
\begin{equation}
{\frac{\partial \delta n^*(\textbf{r},t)}{dt}}=  [\nabla n^{eq}(r)]
\cdot {\bf v}(t) -\int_0^t dt' \int d^3 r'
D^*(\textbf{r},\textbf{r}';t-t')\delta n^*(\textbf{r}',t') +
f(\textbf{r},t), \label{eq4}
\end{equation}
with $D^*(\textbf{r},\textbf{r}';t)$ being the elements of the ``matrix" $D^*(t)\equiv [L_{n n}(t)(\chi^{-1})_{nn}]$. The first term on the right side of this equation is a linearized
streaming term and $f(\textbf{r},t)$ is a fluctuating term, related
to $D^*(\textbf{r},\textbf{r}';t)$ by $\langle f(\textbf{r},t)f(\textbf{r}',t')\rangle= \int
d^3r'' D^*(\textbf{r},\textbf{r}'';t-t')  \chi_{nn}
(\textbf{r}'',\textbf{r}')$.

Formally solving Eq.\ (\ref{eq4}) and substituting the solution for
$\delta n^*(\textbf{r},t)$ in Eq.\ (\ref{eq3}), leads again to the generalized Langevin equation in Eq. (\ref{eq6}) with the time-dependent friction tensor $\stackrel{\leftrightarrow }{\Delta
\zeta (t) }$ of Eq. (\ref{deltazetaconf}) given more concretely by
\begin{equation}
\Delta \stackrel{\leftrightarrow }{\zeta} (t)= -\int d^3 r \int d^3
r' [\nabla u(r)] G^*(\textbf{r},\textbf{r}';t)[\nabla '
n^{eq}(r')] , \label{eq7}
\end{equation}
where $G ^* (\textbf{r},\textbf{r}';t)$ is the propagator, or
Green's function, of Eq.\ (\ref{eq4}),
i.e., it solves the equation
\begin{equation}
{\frac{\partial G ^*(\textbf{r},\textbf{r}';t)}{dt}}= -\int_0^t
dt' \int d^3 r'' D^*(\textbf{r},\textbf{r}'';t-t')G
^*(\textbf{r}'',\textbf{r}'; t') , \label{eq8}
\end{equation}
with initial value $G ^*(\textbf{r},\textbf{r}';t=0)= \delta
(\textbf{r}-\textbf{r}')$. Notice that, since the initial value
$\delta n^*(\textbf{r},t=0)$ is statistically independent of ${\bf
v}(t)$ and $f(\textbf{r},t)$,  the density-density
time-correlation function $\chi^*(\textbf{r},\textbf{r}';t) \equiv
\langle\delta n^*(\textbf{r},t) \delta n^*(\textbf{r}',0)\rangle$, which is the
van Hove function of the particles surrounding the tracer particle,
and observed from the tracer particle's reference frame, is also a
solution of the same equation with initial value
$\chi^*(\textbf{r},\textbf{r}';t=0)=\chi_{nn} (\textbf{r},\textbf{r}')$.

To simplify the notation, let us re-write Eq.\ (\ref{eq7}) as $\Delta
\stackrel{\leftrightarrow }{\zeta} (t)= - [\nabla u^{\dag}] \cdot
G ^*(t)\cdot[\nabla n^{eq}] $, where the convolution $\int d^3
r''A(\textbf{r},\textbf{r}'')B(\textbf{r}'',\textbf{r}')$ between
two arbitrary functions $A$ and $B$ is written as the inner product
$A\cdot B$, and similarly with (column) ``vectors" such as $u$ and
$n^{eq}$. In this notation, the dagger means transpose. With this
notation,  Wertheim-Lovett's relation  reads $[\nabla
n^{eq}] = -\beta \chi_{nn} \cdot [\nabla u]$. With this relation, and the definition of the
the inverse matrix $\chi_{nn}^{-1}$ by the equation $\chi_{nn}^{-1}\chi_{nn} = I$, with $I$
being the unit matrix ($I(\textbf{r},\textbf{r}')\equiv \delta
(\textbf{r}-\textbf{r}')$ = Dirac's delta
function), one can write Eq.\ (\ref{eq7}) in a variety of
different but equivalent manners. In particular, we will
employ the following:

\begin{equation}
\Delta \stackrel{\leftrightarrow }{\zeta} (t)= k_BT [\nabla
n^{eq\dag}]\cdot \chi_{nn}^{-1} \cdot \chi^*(t) \cdot \chi_{nn}^{-1} \cdot
[\nabla n^{eq}], \label{eq9}
\end{equation}

\noindent
where we have used the fact that the van Hove function $\chi^*(t)$ can
be written as $\chi^*(t)=G^*(t)\cdot\chi_{nn}$.

Let us now notice that for spherical particles $\Delta \stackrel {\leftrightarrow}{\zeta} (t)$ is isotropic and diagonal,
\begin{equation}
\Delta \stackrel {\leftrightarrow}{\zeta} (t)=  \stackrel {\leftrightarrow}{I}\Delta \zeta (t),  \label{isotropydzeta}
\end{equation}
so that we only have to calculate the scalar time-dependent friction function $ \Delta \zeta (t)$. The exact expressions for $\Delta \zeta (t)$ can then be given a more concrete and tractable appearance if some approximations are introduced, related to the
general properties of the functions $\chi^*(\textbf{r},\textbf{r}';t)$
and $\chi_{nn} (\textbf{r},\textbf{r}')$. The latter is just the
two-particle distribution function of the colloidal particles
surrounding the tracer particle, but subjected to the ``external"
field $u(r)$ exerted by this tracer particle. Thus, it is
effectively a three-particle correlation function. Only if one
ignores the effects of such ``external" field, one can write $\chi_{nn}
(\textbf{r},\textbf{r}') = \chi_{nn} (|\textbf{r}-\textbf{r}'|)\equiv
n\delta(\textbf{r}-\textbf{r}')+n^2[g(|\textbf{r}-\textbf{r}'|)-1]$.
Similarly, we may also approximate $\chi^*(\textbf{r},\textbf{r}';t)$
by $ \chi^*(|\textbf{r}-\textbf{r}'|;t)$. This is referred to as the
\emph{``homogeneous fluid approximation"}  \cite{faraday}, which then allows us to
write
\begin{equation}
\chi_{nn}(\textbf{r},\textbf{r}';t)=(1/2\pi)^3\int d^3k \exp[
{i\textbf{k}\cdot\textbf{r}}] nS(k)
\end{equation}
and
\begin{equation}
\chi^*(\textbf{r},\textbf{r}';t)=(1/2\pi)^3\int d^3k \exp[
{i\textbf{k}\cdot\textbf{r}}] nF^*(k,t),
\end{equation}
with
\begin{equation}
F^*(k,t)\equiv\frac{1}{N}
\langle\sum_{i,j}^N \exp[i\textbf{k}\cdot[\textbf{r}_i(t)-\textbf{r}_j(0)]]
\rangle.
\label{fstardkt}
\end{equation}
and
\begin{equation}
S(k) = F^*(k,t=0).
\end{equation}
Using these expressions in Eq. (\ref{eq9}), along with the fact that $\nabla n^{eq}(\textbf{r})= n \nabla g(r)= n \nabla h(r) $ (so that its FT is $i\textbf{k}nh(k)=i\textbf{k}[S(k)-1]$), we have that Eq. (\ref{eq9}) becomes
\begin{equation}
\Delta \zeta (t) =\frac{k_BT}{3\left( 2\pi \right) ^{3}n}\int d
{\bf k}\left[\frac{ k[S(k)-1]}{S(k)}\right] ^{2}F^*(k,t).
\label{dzdt00}
\end{equation}

The function $F^*(k,t)$ in this equation is just the intermediate scattering function, but the asterisk indicates that the position vectors $\textbf{r}_i(t)$
and $\textbf{r}_j(0)$ have the origin in the center of the tracer
particle. Denoting by $\textbf{x}_T(t)$ the position of the tracer
particle referred to a laboratory-fixed reference frame, we may
re-write
\begin{equation}
F^*(k,t)\equiv \langle \
\left[\frac{1}{N}\sum_{i,j}^N \exp
(i\textbf{k}\cdot[\textbf{x}_i(t)-\textbf{x}_j(0)])\right]\cdot\left[\exp
(i\textbf{k}\cdot[\textbf{x}_T(t)-\textbf{x}_T(0)])\right]\ \rangle,
\label{decouplingapprox}
\end{equation}
where $\textbf{r}_i(t)$ is the position of the $i$th particle in
the fixed reference frame. Approximating the average of the product
in this expression by the product of the averages, leads to
\begin{equation}
F^*(k,t) \approx F(k,t)F_S(k,t),
\end{equation}
where $F_S(k,t)\equiv \langle \exp
\{i\textbf{k}\cdot[\textbf{x}_T(t)-\textbf{x}_T(0)]\} \rangle$ is the \emph{self} ISF. This is referred to as the \emph{decoupling approximation}  \cite{faraday}. Thus, from the exact result in Eq.\ (\ref{eq9}) above, plus the
introduction of the two approximations just described, we finally arrive at the following general but approximate expression for the time-dependent friction function $\Delta \zeta (t)$, \begin{equation}
\Delta \zeta (t) =\frac{k_BT}{3\left( 2\pi \right) ^{3}n}\int d
{\bf k}\left[\frac{ k[S(k)-1]}{S(k)}\right] ^{2}F(k,t)F_S(k,t).
\label{dzdt0p}
\end{equation}
This expression is reminiscent of the corresponding mode coupling theory (MCT) result \cite{goetze1}. Its derivation above, however, follows a completely different conceptual route. In the following section we discuss important implications of our results above.

\section{Long-time dynamic equivalence.}\label{sectionV}

The main result of the previous sections is, of course, the generalized Langevin equation (Eq. (\ref{eq6})) describing the ballistic to diffusive crossover of the Brownian motion of individual tracer particles in an atomic liquid. Taking into account the isotropy of the configurational time-dependent friction function (Eq. (\ref{isotropydzeta})), this stochastic equation reads
\begin{equation}
M{\frac{d{\bf v}(t)}{dt}}= -\zeta^0 {\bf v}(t)+{\bf f}
^0(t)- \int_0^t dt' {\Delta
\zeta(t}-t')  {\bf v}(t')+ {\bf F} (t). \label{eq6p}
\end{equation}
The configurational effects of the interparticle interactions is embodied in the time-dependent friction function $\Delta \zeta (t) $, which Eq. (\ref{dzdt0p}) writes in terms of the ISFs $F(k,t)$ and $F_S(k,t)$.
Thus, the full analysis of this stochastic equation requires in principle the previous determination of these more complex dynamic properties. Some important implications, however, can be drawn without a detailed knowledge of  $\Delta \zeta (t) $.

The most remarkable conclusion is that the Brownian motion of individual tracer particles in atomic and colloidal  liquids is described by the same mathematical model, namely, the generalized Langevin equation derived here for atomic systems (Eq. (\ref{eq6p}) with  Eq.  (\ref{dzdt0p})), and  the GLE derived in Ref. \cite{faraday} for colloidal fluids (Eq. (\ref{gletracerdif}) with  Eq.  (\ref{dzdt0})). According to this formal mathematical similarity, the properties that describe the tracer's random motion in atomic and in colloidal liquids, such as the mean squared displacement $W(t)$, should collapse onto each other when expressed in dimensionless units that absorb the mass $M$ and the short-time friction coefficient ($\zeta ^{(s)}$ or $\zeta^0$).

To see this, let us first notice that from Eq. (\ref{eq6p}) one can write the velocity autocorrelation function $V(t)$ in terms of  $\Delta \zeta (t) $, in Laplace space, as
\begin{equation}
V(z)=\frac{\frac{k_BT}{M}}{z+\frac{\zeta_S}{M}+\frac{\Delta \zeta
(z)}{M}}, \label{vdzinter}
\end{equation}
with the friction coefficient $\zeta_S$ representing either the kinetic friction coefficient $\zeta^0$ in atomic fluids or the solvent friction coefficient $\zeta ^{(s)}$ in colloidal liquids,
\begin{equation}
\zeta_S = \Big\{
\begin{array}{cccc}
\zeta^0 & & \textrm{(for atomic systems)}   \\
\zeta^{(s)} & & \textrm{(for Brownian systems)}.
\end{array}
\end{equation}
Using this result in the exact relationship in Eq. (\ref{wdzvdz}) one
can derive the
following integro-differential equation for $W(t)$,
\begin{equation}
\tau_S\frac{dW(t)}{dt} +W(t) =D_St-\int_0^t \left[\frac{\Delta \zeta
(t-t')}{\zeta_S}\right]W(t')dt', \label{wdtintdifec}
\end{equation}
where
\begin{equation}
\tau_S\equiv M/\zeta_S \label{taub}
\end{equation}
is the crossover timescale from ballistic to diffusive motion (and which  equals the mean free time $\tau_0$ only in atomic liquids), and where the short-time self-diffusion coefficient $D_S$ is defined by Einstein's relation,
\begin{equation}
D_S\equiv k_BT/\zeta_S. \label{einsteinshort}
\end{equation}
Thus, $D_S=D^0$ is given by the kinetic theoretical result in Eq. (\ref{dkinetictheory}) only for atomic liquids.

For both, atomic and Brownian tracers, in the absence of interactions $\Delta \zeta
(t)$ vanishes and Eq. (\ref{wdtintdifec}) becomes Eq. (\ref{wdtintdifec0}), discussed in Sec. \ref{sectionII} for atomic liquids, and whose solution is given by Eq. (\ref{wdt0}). In the presence of interactions, however, $\Delta \zeta (t)\ne0$,
but at very short times ($t\ll
\tau_S$) the solution of  Eq. (\ref{wdtintdifec}) is still identical to that of a freely-flying particle,
i.e., the short-time asymptotic expression for the MSD is also given by $W(t)
\approx \frac{1}{2} v_0^2 t^2$, as illustrated by the molecular dynamics simulations in  Fig. \ref{fig1}.

In the opposite regime, $t\gg \tau_S$, the interparticle interactions change the long-time asymptotic limit of $W(t)$ from its free-diffusion value $W(t) \approx D^0t$ to the new value $W(t) \approx D_Lt$, which defines the long-time self-diffusion coefficient $D_L$.  In this regime, the convolution $\int_0^t \Delta \zeta
(t-t')W(t')dt'$ on the right side Eq. (\ref{wdtintdifec}) can be approximated by its Markov limit $ [\int_0^\infty \Delta \zeta
(t')dt']W(t) = \Delta \zeta W(t)$, so that  Eq. (\ref{wdtintdifec}) reads
\begin{equation}
\tau_S\frac{dW(t)}{dt} +\frac{W(t)}{D^*} =D_St, \label{wdtintdifecMarkov}
\end{equation}
where
\begin{equation}
D^*\equiv \frac{D_L}{D_S} = \frac{1}{1+\Delta \zeta/\zeta_S},
\label{dstarr}
\end{equation}
with the constant $\Delta \zeta $ defined as $\Delta \zeta \equiv \int_0^\infty \Delta \zeta(t)dt.$
This equation is precisely Eq. (\ref{wdtintdifec1}) of Sec. \ref{sectionII}, whose analytic solution in Eq. (\ref{wdtstar1}) was shown in Fig. 1 to provide a simple interpolation between the short- and long-time limits of the molecular dynamics simulation data for the MSD of the HS liquid throughout its thermodynamically stable liquid regime, $0\le \phi \lesssim 0.5$.

To continue the discussion of the dynamic equivalence between atomic and Brownian liquids, let us define in both cases a length $l_S$ in terms of $\tau_S$ and $D_S$ as
\begin{equation}
l_S^2\equiv D_S\tau_S. \label{ls}
\end{equation}
For atomic liquids $l_S$ is identical to the mean free path $l_0$, but not for colloidal systems, for which $l_S$ is only given by this equation (together with Eqs. (\ref{taub}) and (\ref{einsteinshort})). We may now use $l_S$ and $\tau_S$ as the units of length and time, respectively, and rewrite Eq. (\ref{wdtintdifec}) in terms of the scaled time $t^*\equiv t/\tau_S$ and the scaled MSD $w(t^*) \equiv \ W(t)/l_S^2 $ as
\begin{equation}
\frac{dw(t^*)}{dt^*} +w(t^*) =t^*-\left(\frac{\tau_S}{\tau_I}\right)\int_0^{t^*} \Delta \zeta^*
(t^*-t^{*'})w(t^{*'})dt^{*'}, \label{wdtintdifecstar}
\end{equation}
where the configurational timescale $\tau_I$, the time it takes a particle to diffuse a mean distance $d\equiv n^{-1/3}$ with a diffusion coefficient $D_S$, is given by
\begin{equation}
\tau_I\equiv d^2/D_S, \label{tauint}
\end{equation}
and in which we have defined the dimensionless function
$\Delta \zeta^* (t)$ as
\begin{equation}
\Delta \zeta^* (t)\equiv \left[\frac{\tau_I  \Delta \zeta
(t)}{\zeta_S}\right] =\frac{1}{3\left( 2\pi \right) ^{3}n^{5/3}}\int d
{\bf k}\left[\frac{ k[S(k)-1]}{S(k)}\right] ^{2}F(k,t)F_S(k,t).
\end{equation}

Let us notice that the purpose of using $l_S$ and $\tau_S$ as the units of length and time is to  focuss on the crossover time-regime from ballistic to diffusive motion. If, instead, we were interested in focussing on the crossover  from free-diffusion to structural relaxation, the best would be to use the mean inter-particle distance $d\equiv n^{-1/3}$ as the unit of length and $\tau_I$ as the time unit, and to rewrite Eq. (\ref{wdtintdifec}) in terms of the scaled time $t^*\equiv t/\tau_I$ and the scaled MSD $w(t^*) \equiv \ W(t)/d^2 $, to read
\begin{equation}
\left(\frac{\tau_S}{\tau_I}\right)\frac{dw(t^*)}{dt^*} +w(t^*) =t^*-\int_0^{t^*} \Delta \zeta^*
(t^*-t^{*'})w(t^{*'})dt^{*'}.  \label{wdtintdifecstardiff}
\end{equation}
For dense atomic liquids (e.g., hard spheres at $\phi \lesssim 0.5$) either choice is perfectly adequate to observe  within the same time window both, the crossover from ballistic to diffusive motion and the crossover from free to correlated motion, as illustrated by the simulation results for $W(t)$ in Fig. 1. The reason is, of course, that in this case  $\tau_S$ and  $\tau_I$ do not greatly differ from each other, and hence, there is no important time-scale separation.

The situation is, however, dramatically different in the corresponding colloidal liquid since, as discussed in Sect. \ref{sectionII}, the ratio $\tau_S/\tau_I$ may be as small as $\tau_S/\tau_I\approx 10^{-8}$. Thus, these two crossover timescales are separated by about 8 decades, and cannot be analyzed in the same time window. In fact, this implies that if we focus on the crossover from ballistic to diffusive motion, as  in Eq. (\ref{wdtintdifecstar}), then the term involving the configurational friction will be completely negligible. In contrast, if we focus on the  crossover from free to correlated motion, as  in Eq.  (\ref{wdtintdifecstardiff}), then it is the inertial term involving the time derivative of the MSD what can be neglected. This is referred to as the overdamped limit, which also amounts to ignoring the inertial term $M[d{\bf v}(t)/dt]$ on the left side of the GLE in Eq. (\ref{eq6p}). Thus, in this limit Eq.
(\ref{wdtintdifecstardiff}) reads
\begin{equation}
w(t^*) =t^*-\int_0^{t^*} \Delta \zeta^*
(t^*-t^{*'})w(t^{*'})dt^{*'}. \label{wdtintdifecstaroverdamped}
\end{equation}
This only changes the true short-time limit $w(t^*) \approx (d/l_S)^2t^{*2}/2$ of the solution of Eq. (\ref{wdtintdifecstardiff})
to $w(t^*) \approx  t^*$, but leaves unaltered the long-time limit
$w(t^*) \approx D^* t^*$. This equation thus describes the diffusive motion of colloidal tracer particles.

It should also be clear, however, that even though in atomic liquids there is not an appreciable timescale separation, Eq. (\ref{wdtintdifecstaroverdamped}) also describes the long-time motion of atomic tracer particles.
Hence, except for the referred short-time
differences, the MSD of an atomic and a colloidal liquid with the
same interactions and the same  $S(k)$ should be indistinguishable
when plotted in terms of these dimensionless units, provided that the intermediate scattering functions $F(k,t)$ and $F_S(k,t)$, which enter in the previous expression for $\Delta \zeta^* (t)$ above, also share a similar long-time scaling property. Thus, we may embark on a study of these dynamic properties to see if they indeed exhibit the desired scalings, or else, we can check directly if the MSD itself exhibits the expected universality of
atomic and colloidal liquids.

Here we have adopted the second approach, and have compared the molecular and the Brownian dynamics simulation data of $W(t)$ for the HS system at the volume fractions $\phi=0.1$ and 0.5.  For this comparison we use the same  molecular dynamics data as in  Fig. \ref{fig1}, now plotted in terms of the dimensionless MSD $ [W(t)/6d^2] $ as a function of the dimensionless time
$ t/\tau_I=[D^0t/d^2]$, and the Brownian dynamics data generated for this comparison using the methodology explained in Ref. \cite{dynamicequivalence}. As we can see from this comparison, for each volume fraction the molecular dynamics and the Brownian dynamics data agree at long times, within a high degree of numerical precision. The short-time difference between the molecular and Brownian dynamics simulation data originates, of course, from the fact that the latter are based on the conventional Ermak and McCammon's Brownian dynamics algorithm \cite{tildesley,ermakmccammon}, in which the ``overdamped" limit is previously taken in the microscopic equations of motion.  Thus, this agreement is also an indirect indication that the intermediate scattering functions $F(k,t)$ and $F_S(k,t)$ must also share similar scaling properties. The analysis of this issue, however, is addressed separately \cite{overdampedatomic}.

\begin{figure}
\begin{center}
\includegraphics[scale=.27]{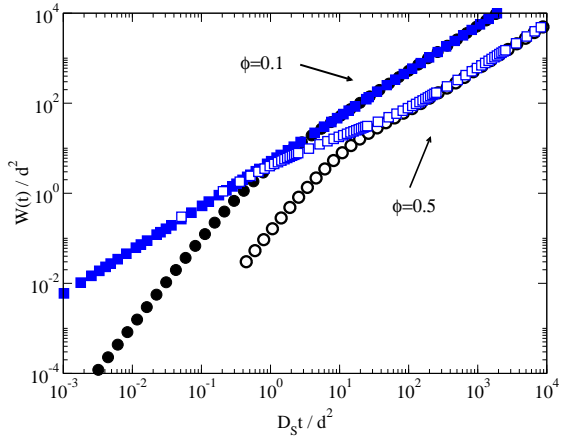}
\caption{Dimensionless mean squared displacement $[W(t)/d^2]$ as a function of the dimensionless time
$[D_S t/d^2]$ for the volume fractions $\phi=0.1$ and 0.5. The squares correspond to Brownian
Dynamics simulations \cite{tildesley,ermakmccammon} generated as described in Ref. \cite{dynamicequivalence}, and the circles to the molecular dynamics simulations of Fig. 1. }
\label{fig2}
\end{center}
\end{figure}

\section{Discussion and summary.}\label{sectionVI}

In summary, in this paper we have explained a simple argument that exposes a dynamic equivalence between the long-time dynamic properties of atomic and colloidal liquids. Such simple arguments were complemented by a more formal fundamental derivation of the generalized Langevin equation for a tracer particle in an atomic liquid, which is the atomic counterpart of the GLE for tracer diffusion derived in Ref. \cite{faraday} for colloidal liquids in the absence of hydrodynamic interactions. The dynamic equivalence suggested by the fact that the GLE for tracer diffusion in both cases has the same mathematical structure, need in reality that other dynamic properties, such as the intermediate scattering functions $F(k,t)$ and $F_S(k,t)$, also share a similar long-time scaling property. For the time being, here we have tested the predicted dynamic equivalence at the level of the mean squared displacement $W(t)$ in the context of a specific model system, namely, the hard sphere liquid, in its dynamic version corresponding to molecular and Brownian dynamics.

Let us state, however, that the present work does not settle the question of the generality of this dynamic equivalence. Instead, it only contributes to stimulate the corresponding discussion. For example, it is important to discuss the manifestation of this dynamic equivalence on properties other than the MSD. As indicated above, verifying that similar scalings are exhibited by the intermediate scattering functions $F(k,t)$ and $F_S(k,t)$ is an issue that must still be addressed  in detail. In fact, our group has already approached this issue within the GLE formalism \cite{overdampedatomic}, and the results turn out to be completely consistent with those of the present paper. The atomic-to-Brownian long-time dynamic equivalence thus seems to be a very robust prediction. The most relevant implications of this dynamic equivalence have been corroborated by the systematic comparisons between molecular and Brownian dynamics simulations of the sort illustrated in this paper. A summary of this analysis has been advanced in a recent brief communication \cite{atombrownequivletter}. Another important issue refers to the actual universality of the atomic-colloidal dynamic equivalence discussed in this paper, since the only validation of these predictions involved a specific model system, namely, the hard sphere fluid.

In this direction, let us mention that the present
colloidal--atomic dynamic correspondence is not restricted to the
hard-sphere fluid, but it actually extends over to systems with soft
repulsive interactions. This is a direct result of combining the
present colloidal--atomic correspondence for the hard sphere system,
with another important scaling rule, which derives from the
principle of dynamic equivalence between soft-sphere and hard-sphere
liquids \cite{dynamicequivalence,soft2}. The extension of this scaling to atomic systems is immediate once
the collision diameter $\sigma$ entering in the expression for
$D^0(n,T)$ in Eq. (\ref{dkinetictheory}) is given a proper
definition \cite{softhardatomiccolloidal} for the soft-sphere potential $u(r)$ considered.

Still another issue refers to the possible limitations of this long-time dynamic equivalence, imposed by the fact  that the present derivation apparently assumed colloidal systems in the absence of hydrodynamic interactions. In reality, however, the validity of this dynamic equivalence should extend over to systems with hydrodynamic interactions, provided that the corresponding effects enter only through the value of the short-time self-diffusion coefficient $D_S$, as suggested in \cite{prlhi}. Other interesting directions along which to question the applicability and universality of this dynamic equivalence refers to the realm of liquid mixtures and to the effects of attractive interactions. The answer to these questions, however, will only come from the comparison between the dynamic properties of atomic and colloidal liquids, similar to that presented here in Fig. 2, or by expanding the theoretical analysis that led us to the present proposal.

\bigskip

ACKNOWLEDGMENTS: The authors are grateful to G. P\'erez-\'Angel and M. Ch\'avez-P\'aez for their advise and assistance with the molecular and Brownian dynamics simulations, and to L. Yeomans-Reyna, for valuable discussions. We also acknowledge the kind hospitality of the Joint Institute for Neutron Sciences (Oak Ridge, TN), where part of this manuscript was written. We are grateful to W.-R. Chen and T. Egami for stimulating discussions.
This work was supported by the Consejo Nacional de
Ciencia y Tecnolog\'{\i}a (CONACYT, M\'{e}xico) through grants 84076 and
132540 and through the Red Tem\'atica de la Materia Condensada Blanda.

\end{document}